\documentclass{emulateapj}
\shorttitle{Si and O in planet hosts}
\shortauthors{Brugamyer et al.}
\usepackage{rotating}
\usepackage{mdwlist}

\begin{document}

\title{Silicon and Oxygen Abundances in Planet-Host Stars}

\author{Erik Brugamyer\altaffilmark{1}, Sarah E. Dodson-Robinson, William D. Cochran, and Christopher Sneden}
\affil{Department of Astronomy and McDonald Observatory, University of Texas at Austin \\
1 University Station C1400, Austin, TX 78712}


\altaffiltext{1}{erikjb@astro.as.utexas.edu}

\begin{abstract}

The positive correlation between planet detection rate and host star iron abundance lends strong support to the core accretion theory of planet formation.  However, iron is not the most significant mass contributor to the cores of giant planets.  Since giant planet cores are thought to grow from silicate grains with icy mantles, the likelihood of gas giant formation should depend heavily on the oxygen and silicon abundance of the planet formation environment.  Here we compare the silicon and oxygen abundances of a set of 76 planet hosts and a control sample of 80 metal-rich stars without any known giant planets.  Our new, independent analysis was conducted using high resolution, high signal-to-noise data obtained at McDonald Observatory.  Because we do not wish to simply reproduce the known planet-metallicity correlation, we have devised a statistical method for matching the underlying [Fe/H] distributions of our two sets of stars.  We find a 99\% probability that planet detection rate depends on the silicon abundance of the host star, over and above the observed planet-metallicity correlation.  We do not detect any such correlation for oxygen.  Our results would thus seem to suggest that grain nucleation, rather than subsequent icy mantle growth, is the important limiting factor in forming giant planets via core accretion.  Based on our results and interpretation, we predict that planet detection should correlate with host star abundance for refractory elements responsible for grain nucleation and that no such trends should exist for the most abundant volatile elements responsible for icy mantle growth.


\end{abstract}


\section{Introduction}
\label{intro}

The tendency for planets to orbit metal-rich stars lends strong support to the core accretion model of planet formation, whereby planets grow through accretion of solid, metal-rich material to form massive cores. Within the context of core accretion (cf. Safronov 1969; Pollack et al. 1996), heavy element abundances are important to the extent that they contribute to the inventory of solid material available for planetesimal formation.  Iron (the typically-used proxy for overall stellar metallicity) is certainly an important component, but there are other significant contributors, especially oxygen, carbon, silicon, magnesium, sulfur, nitrogen and aluminum.

Oxygen is thought to be the single most important contributor to the mass of giant planets, primarily via water ice accreted beyond the snow line of the disk (Hayashi 1981, Weidenshilling 1977) and, to a lesser extent, through the oxides of Si, Mg, Ca, and Al.  Carbon, via heavy organic compounds, is probably the second most important mass contributor (Lodders 2004), followed by silicon.  These elements often demonstrate different abundance patterns relative to iron.  Robinson et al. (2006) reported relative silicon and nickel enrichment in planet hosts and Fuhrmann \& Bernkopf (2008) have reported enhancements in alpha-capture elements.  Thus, iron is likely not an ideal proxy for measuring the abundances of material used to build planet cores.

Previous tests of how individual elements contribute to planet formation have focused on the possibility that planet hosts are chemically peculiar stars with abundance ratios that differ from typical Population~I stars.  If planet hosts are chemically peculiar, the slopes of [X/Fe]~vs.~[Fe/H] among them should be distinct from what Galactic chemical enrichment models (e.g. Timmes et al. 1995) predict.  Bodaghee et al. (2003) found no such differences in their sample for alpha- and iron-peak elements.  They observe no difference in the overall trends of [X/Fe] between planet hosts and their volume-limited sample of stars without any known planetary-mass companions.  Based on their results, stars with planets appear to be indistinguishable from other field stars, and seem to simply lie on the high-metallicity end of otherwise ``normal'' stellar distributions.

Given the metal-rich nature of planet-hosting stars, a pressing need when further exploring the planet-formation importance of individual elements is to carefully match the underlying iron distributions of planet-host and control samples.  Previous planet-host studies (e.g. Fischer \& Valenti 2005, Neves et al. 2009) have found that, in addition to iron, the abundances of various other metals are enhanced in these stars compared to stars with no known planets.  This comes as no surprise, however, given the known positive correlation between host star iron abundance and planet detection rate.  Since planet-hosts are found to have higher overall iron content (which serves as a proxy for overall metallicity) compared to non-hosts, they are indeed expected to have a higher content of other metals as well.  An ideal study would consist of an arbitrarily large number of hosts and non-hosts, such that the samples could be divided into arbitrarily small [Fe/H] bins and still have a statistically significant number of hosts and control stars in each bin.  It would then be trivial to determine if, \emph{at a given [Fe/H]}, a difference existed between the two groups of stars in the average abundances of elements important for planet formation.  In the absence of such an ideal sample, we have devised a statistical method for matching the underlying iron distributions.

The present analysis is aimed at examining the most abundant heavy elements important for planet formation.  We have chosen to focus first on silicon and oxygen.  Our hypothesis is that if core-accretion is responsible for the majority of known giant planets, then for a given [Fe/H] their stellar hosts should show enhancements in silicon and oxygen relative to iron.  We therefore wish to determine whether there is a statistically significant difference in the silicon and oxygen abundance distributions of stars with planets, compared to those without any known giant planets.

\section{Observations and Data Reduction}
\label{observations}

For this study, we selected 76 FGK dwarf and sub-giant host stars and 80 non-host stars for comparison.  The data were obtained between July 1998 and March 2010 on the 2.7-meter Harlan J. Smith telescope and the Hobby-Eberly telescope (``HET'') at McDonald Observatory.  Our program stars span the following ranges: $-0.67~\leq~[Fe/H]~\leq~+0.54$; $4935~\leq~T_{\rm eff}~\leq~6250$~(K); $3.15~\leq~\log \: g~\leq~4.63$; and $0.54~\leq~v_{\rm mic}~\leq~1.53$~(km~s\textsuperscript{-1}), where $v_{\rm mic}$ represents the microturbulent velocity.

\subsection{Non-host stars}

All non-host stars were selected from the $\backsim$300 stars being monitored as part of the McDonald Observatory Planetary Search program (hereafter ``MOPS''; see Wittenmyer et al. 2006 for a description of the program and detection limits) on the 2.7m telescope.  For these, we used template spectra taken without the iodine cell in the optical path.  Using the current instrumental configuration (``Phase III''; begun in 1998), the program achieves routine internal precision of 6--9~m~s\textsuperscript{-1}.  With a monitoring baseline of over 10 years, we can thus exclude roughly Jupiter-mass companions out to 5~AU, or roughly Neptune-mass companions out to 1~AU, around these stars.

Since planet-hosting stars tend to have higher overall metallicity, we built our non-host sample by choosing the most metal-rich stars available from the MOPS program.  This was done by cross-referencing the MOPS list with available metallicity references from the SIMBAD Astronomical Database\footnote{http://simbad.u-strasbg.fr/simbad/} and the NASA/IPAC/NExScI Star and Exoplanet Database (``NStED'')\footnote{http://nsted.ipac.caltech.edu/} and choosing non-host stars in such a manner that the final overall metallicity distributions of our host and non-host samples were as similar as possible.  Note that we attempt to statistically control for imperfect matching of the [Fe/H] distributions later in our analysis.

\subsection{Host stars}

Our host stars were selected in a statistically haphazard manner, as follows.  Data for twenty-six of our planet-host stars came from the MOPS program (as in the case of the non-host stars), by selecting only those MOPS host stars with data having a signal-to-noise ratio $\geq$~100.  The remaining 50 hosts were observed independently, using both the 2.7m telescope (in the same instrumental setup as the MOPS program) and the HET.  For these, we selected the brightest objects with confirmed planetary companions in the literature that were available for observation from McDonald Observatory during our supplemental observing runs from December 2009 to March 2010.  Fifty-six of our host stars were ultimately observed with the 2.7m telescope, and 20 with the HET.

\subsection{Instruments}

For the 2.7m telescope, we utilized the Tull Coud\'{e} Spectrometer (Tull et al. 1994).  This cross-dispersed echelle spectrograph uses a 2048x2048 Tektronix CCD with 24~$\mu$m pixels and our configuration uses the ``E2'' grating with 52.67~groove~mm\textsuperscript{-1}.  With a 1.2~arcsec slit, we achieve a resolving power (=$\lambda/\Delta\lambda$) of R=60,000 in this configuration.  The wavelength coverage extends from 3750~\r{A} to 10,200~\r{A}.  Coverage is complete from the blue end to 5691~\r{A}, with increasingly large inter-order gaps thereafter.

For the HET, we utilized the fiber-fed High Resolution Spectrograph (Tull 1998).  The spectrograph uses a mosaic of two 2048x4102 Marconi Applied Technologies (now E2V Technologies) CCDs with 15~$\mu$m pixels and a grating with 316~groove~mm\textsuperscript{-1}.  Using a 2.0~arcsec fiber, we achieve a resolving power of R=60,000 with this instrument.  The wavelength coverage extends from 4090~\r{A} to 7875~\r{A}.  Coverage is complete except for the range 5930~\r{A} to 6012~\r{A}, corresponding to the gap between the two CCDs.  The signal-to-noise ratio of our 2.7m and HET data range from $\backsim$100-500.

\subsection{Data reduction}

The data were reduced using standard routines within the \emph{echelle} and \emph{onedspec} packages of the Image Reduction and Analysis Facility (IRAF).  The process included overscan trimming, bias frame subtraction, removal of scattered light, flat field division, extraction of the orders and wavelength calibration using a Th-Ar calibration lamp spectrum.  We then manually removed any cosmic rays that IRAF's interpolation routines were unable to handle.  The final steps involved dividing out the blaze function, normalizing the continuum and combining orders.

\section{Measuring Abundances}
\label{moog}

The results of our stellar parameter and abundance determinations are listed in Table 1 (for planet-host stars) and Table 2 (for non-host stars).  Our process of determining stellar parameters and abundances involved the following steps:

\begin{enumerate}

\item We constructed a list of neutral and singly-ionized iron lines.
\item We obtained a spectrum of the Sun.
\item We measured the equivalent widths of each of our selected iron lines in the solar spectrum.
\item We used these measurements to determine the solar parameters (effective temperature, surface gravity, microturbulent velocity and overall metallicity).
\item With our final solar model atmosphere we then determined the silicon and oxygen abundances using spectral synthesis.
\item Steps 2-5 were repeated for each of our target stars.

\end{enumerate}

The process is explained in further detail in the following subsections.

\subsection{Atmospheric parameters and iron abundances}

All stellar parameters and abundances were determined using MOOG\footnote{available at http://www.as.utexas.edu/$\backsim$chris/moog.html} (Sneden 1973) -- a local thermodynamic equilibrium (LTE) line analysis and spectrum synthesis code -- and a grid of Kurucz (1993a) ATLAS9 model atmospheres.  We constrained the stellar parameters of our targets using a carefully-selected list of 65 isolated, unblended neutral iron lines and 22 singly-ionized iron lines, spanning a wide range in excitation potentials and equivalent widths.  The equivalent widths of each of these lines was measured in our program stars using an Interactive Data Language (IDL) routine written exclusively for this purpose (cf. Roederer et al. 2010).  The program fits either a gaussian or voigt profile to each line, and allows for manual adjustment of the continuum.  The program output is a list of equivalent widths for use with MOOG.

MOOG force-fits abundances to match the measured equivalent widths for each line, using the input atomic line parameters (wavelength, excitation potential and oscillator strength).  Our Fe I line parameters, including oscillator strengths (or ``log~\emph{gf}'' values), were taken from the National Institute of Standards and Technology (NIST) Atomic Spectra Database\footnote{http://www.nist.gov/physlab/data/asd.cfm}, supplemented with values from O'Brian et al.~(1991).  For the NIST values, we used only those lines with a log~\emph{gf} accuracy grade of ``D'' or better (i.e. we excluded the lowest-quality ``E'' data from our analysis).  See Table 3 for the full list of neutral iron lines used in our analysis.  Our Fe~II parameter values were taken from Mel\'{e}ndez \& Barbuy (2009) and are listed in Table 4.

\subsection{Calibration using the solar spectrum}

We began the analysis by measuring iron line equivalent widths in a spectrum of the daytime sky, taken through the solar port of the 2.7m telescope.  With these measurements in hand, we then used MOOG to constrain the effective temperature by eliminating any trend of iron abundance with excitation potential (i.e. by assuming excitation equilibrium).  The microturbulent velocity was determined by eliminating any trend of abundance with reduced equivalent width (=~EW/$\lambda$).  The surface gravity was constrained by forcing the derived abundance using singly-ionized iron to match that of neutral iron (i.e. by assuming ionization equilibrium).  During this process we rejected any lines that did not give  ``solar-like'' parameters (most likely due to inaccurate oscillator strengths), leaving us with our final list of 65 Fe I and 22 Fe II lines, which was used to constrain the stellar parameters of all our target stars.  Once these requirements were met, we used the resulting stellar parameters to construct a final model atmosphere and used this model to derive an average iron abundance.  Figure~1 shows the results of these measurements, by plotting the derived solar Fe abundance from each line as a function of reduced equivalent width (top panel), or as a function of excitation potential (bottom panel).  Our derived stellar parameters and iron abundance (used as a proxy in the model atmosphere for the overall metallicity) agree well with canonical values.  From our fiducial solar spectrum, we derive $T_{\rm eff}$~=~5780~$\pm$~70~K, $\log \: g$~=~4.50~$\pm$~0.08~dex, and $v_{\rm mic}$~=~1.16~$\pm$~0.04~km~s$^{-1}$.  We measure an iron abundance of $\log \: \epsilon$(Fe)~=~7.52~$\pm~$0.04.  Here we are using the normal notation where $\log \: \epsilon$(X)~=~12.00~+~log~N(X)/N(H), so that $\log \: \epsilon$~=~12.00 for hydrogen.

We wish to stress that our derived iron abundances are based on one-dimensional, hydrostatic model atmospheres and the assumption of LTE.  As such, they are likely subject to various uncertainties, including surface inhomogeneities and non-LTE effects (see Asplund 2005 for a thorough discussion of these effects); and perhaps even the effects of magnetic fields (Fabbian et al. 2010).  However, non-LTE calculations predict relatively small effects for Sun-like stars, with larger effects seen at higher temperatures and lower surface gravities and metallicities.  For our metal-rich sample of FGK dwarf and sub-giant stars, we expect non-LTE effects to be minimized.  Furthermore, since our targets are sun-like stars, we expect any non-LTE effects to calibrate out (at least partially) when calculating a differential abundance with respect to the Sun.

The process described above was repeated for each star in our sample in order to determine stellar parameters and iron abundance.  We then took the difference, on a line-by-line basis, of the derived iron abundance in the star and that of the solar port spectrum.  Note that by calculating an average iron content difference based on a line-by-line analysis, uncertainties in the log~\emph{gf} values are removed from the differential iron abundance.  Our derived iron abundances are thus quoted relative to the Sun in all cases.

\subsection{Spectral synthesis of silicon and oxygen lines}

With estimates of the stellar parameters in hand, we then determined the silicon and oxygen abundances.  For these, we synthesized a small portion of the spectrum around each absorption feature considered.  The program then varies the abundance of the species until the best fit (the minimum residual) to the observed spectrum is found.  For Si we used six neutral lines between 5645~\r{A} and 5793~\r{A} in our analysis, listed in Table~5.  The initial line lists used to construct the synthetic spectra were taken from the Kurucz~(1993b) atomic linelist, and oscillator strengths were then adjusted where necessary to match our solar port spectrum.  Figure~2 shows an example of the synthesis process for the Si~I line at 5708~\r{A}.  We find an average Si abundance of $\log \: \epsilon$(Si)~=~7.58~$\pm$~0.03 from these six lines for the Sun.  This value is somewhat higher than the $\log \: \epsilon$(Si)~=~7.51~$\pm$~0.03 reported in the recent work of Asplund et al. (2009); however, our stellar [Si/Fe] values are independent of the exact derived solar silicon abundance since our stellar abundances are ultimately calculated differentially with respect to the Sun.  Note that we have chosen to ignore non-LTE effects in our Si I analysis, as these effects are thought to be quite small -- on the order of 0.01 dex -- in the Sun (Wedemeyer 2001, Shi et al. 2008).

For oxygen, we used the allowed transition triplet at 7771~\r{A}, 7774~\r{A} and 7775~\r{A}.  These lines have high excitation potentials (9.15~eV) and are known to be formed under conditions that depart significantly from the LTE approximation (cf. Kiselman 1993, 2001).  These non-LTE effects are appreciable -- on the order of a few tenths of a dex (LTE assumptions always result in an overestimation of the abundance derived from the oxygen triplet) -- and are sensitive to stellar atmospheric parameters (temperature, gravity and overall metallicity) and to the efficiency of collisions with hydrogen that is adopted in the non-LTE calculations (Takeda \& Honda 2005, Fabbian et al. 2009).  We have chosen to apply the non-LTE corrections of Ram\'{i}rez, Allende Prieto \& Lambert (2007) to our derived oxygen abundances.  (Note that they chose to ignore inelastic collisions with neutral hydrogen atoms, as these collisions are expected to play a small role at solar metallicities.)  We first synthesized each line of the triplet individually, and then applied the corrections on a line-by-line basis.  For the Sun, we obtained an average non-LTE correction of -0.13~dex, yielding an oxygen abundance of $\log \: \epsilon$(O)~=~8.70~$\pm$~0.04.  This is in good agreement with the $\log \: \epsilon$(O)~=~8.69 value reported by Asplund et al. (2009); although, as with silicon, our differential stellar [O/Fe] values are independent of the exact derived solar oxygen abundance value.  The average $\log \: \epsilon$(O) non-LTE correction for all our targets was \mbox{-0.16~dex}, with a range from \mbox{-0.39~dex} to \mbox{-0.04~dex}.  The application of the corrections reduced the scatter in our [O/Fe] measurements by approximately 20\%.

The final step in the process of determining silicon and oxygen abundances was to calculate the difference, on a line-by-line basis, between our target stars and the Sun.  Our quoted Si and O abundances are therefore differential with respect to the Sun in all cases.


\section{Measurement Repeatability and Uncertainties}
\label{errors}

In an effort to characterize the random errors in our atmospheric parameter determinations, we obtained and analyzed 22 separate observations of the field dwarfs 47~UMa (a G1~V star) and 70~Vir (G4~V) on the 2.7m telescope, and 20 separate observations of 70~Vir with the HET.  These observations were made using the same instruments and configurations as our program stars, and were subjected to identical analysis methods.  Figures~3 and 4 show histograms of the derived [Fe/H] and T$_{\rm eff}$ for the two sets of observations of 70~Vir.  The maximum standard deviations for these measurements are 0.01~dex for [Fe/H]; 10~K for effective temperature; 0.03~dex for $\log \:$~\emph{g}; and 0.04~km~s\textsuperscript{-1} for microturbulent velocity ($v_{\rm mic}$).  Our analysis of 47~UMa yielded similar results.  These represent our internal ``repeatability'' precisions.  The small offsets of 20~K in the mean derived effective temperature and 0.01~dex in the mean derived [Fe/H] are most likely due to the differing number of pixels per resolution element of the two instruments.  Given that these offsets are both a factor of a few smaller than our total error budget, we have not attempted to correct for these minimal differences.

To get an estimate of the accuracy of our measurements, we compared our results to other work in the literature.  Thirty of our 76 planet-host stars also appear in the Fischer \& Valenti (2005) dataset, and 31 of the remaining 46 have stellar parameters listed from at least one source in SIMBAD and/or NStED.  Sixty-two of our 80 non-host stars have stellar parameters listed from at least one source in the online databases.  For these, we averaged the difference between our results and those of Fischer \& Valenti or the databases, yielding standard deviations of 0.06~dex for [Fe/H]; 70~K for effective temperature; and 0.08~dex for log \emph{g}.  Note that microturbulent velocities are not typically reported in the literature.

In order to characterize our systematic errors, we then performed an analysis of the sensitivity of our derived abundances to variations in the stellar atmospheric parameters.  We chose 16~Cyg~B -- a G3~V dwarf -- for this analysis, as its stellar parameters are typical of our sample.  After varying the parameters by the amounts discussed in the previous paragraph, we find systematic uncertainties of $\pm$~0.02~dex in silicon and $\pm$~0.11~dex in oxygen.

The precisions of our Si and O measurements are approximately 0.03~dex and 0.04~dex, respectively.  We estimated these by simply averaging the standard deviations of the derived abundances for the six Si lines and three O lines in all our program stars.  Adding these random uncertainties and the systematic uncertainties in quadrature yields final errors of $\pm$~0.04~dex for silicon and $\pm$~0.12~dex for oxygen.  Since all measurements were made relative to the Sun, we stress that we have not attempted to determine the absolute abundances of Si or O; rather, we simply wish to rank our targets from least to most silicon/oxygen-rich.

\section{Statistical Methods}
\label{statistics}

Figures 5 and 6 show scatter plots of our derived [Si/Fe] and [O/Fe] abundances, as a function of [Fe/H].  In both cases, we recover the overall distribution expected for field stars (e.g. Timmes et al. 1995), whereby Si and O are overabundant relative to Fe for more metal-poor stars.  The slope of [Si/Fe] flattens out around [Fe/H] of zero, while the slope of [O/Fe] turns further negative at supra-solar metallicities.  Thus, we see that the distributions follow a sequence constrained by the Galactic chemical enrichment history.  For metallicity bins already well-populated in our sample ([Fe/H] of -0.2 to +0.4~dex), we observe that our host stars tend to be enriched in Si compared to non-host stars.  Indeed, two-thirds of our host stars lie at or above [Si/Fe] of zero; while three-fourths of our non-host stars lie at or below [Si/Fe] of zero.  We observe no such tendency for oxygen.

In order to quantify any potential differences in the Si and O distributions of stars with and without planets, the fact that planet-hosting stars tend to have higher overall metal content than typical nearby stars needs to be taken into account.  The more metal-rich nature of stars hosting giant planets now seems well-established (Gonzalez 1997, 1998, 1999; Santos et al. 2005; Fischer \& Valenti 2005).  This means that the locus of typical giant-planet-hosting stars is shifted towards higher overall [Fe/H] compared to the locus of typical local field stars.  Figure 7 is a cartoon depicting the situation.  Stars with and without planets appear to follow the same general Galactic chemical evolution trend (Bodaghee et al. 2003, Santos et a. 2005), but because this trend is not flat, the expected average [Si/Fe] of the two groups of stars is different, for reasons having nothing to do with planets.  Since any difference in oxygen and silicon that is related to planets would be a small effect, it is imperative that the overall [Fe/H] distribution of any studied planet-host sample match that of the control sample of field stars.  Ideally, this could be accomplished by constructing arbitrarily large samples of stars, making it possible to separate the planet-host and control samples into arbitrarily small [Fe/H] bins and still leave a statistically significant number of targets in each bin.  In such an ideal case, determining whether a significant difference in the silicon and oxygen content between the two samples is present would be trivial.  Absent an infinite data set, another possibility is to select matching samples a priori -- that is, to ensure samples are constructed such that every planet-host has a matching control star at the exact same [Fe/H].  As described in section 2, we chose non-host stars for our analysis in such a manner that the distributions of [Fe/H] for our host and non-host samples were as similar as possible.  Perfect matching proved impossible, however, given the finite number of non-hosts in the MOPS parent sample, as well as the uncertainties in determining stellar parameters -- targets selected beforehand for a specific [Fe/H] often ended up at a slightly different [Fe/H] after analysis.  We describe our approach to the problem of matching metallicity distributions -- and of compensating for imperfect matching -- below.


Although our samples contain large numbers of targets, we nevertheless do not have enough data to adequately bin the samples by [Fe/H] and still have a statistically significant number of stars in each bin.  We therefore require a statistical method for quantifying the difference, or lack thereof, between the distributions of [Si/Fe] and [O/Fe] in our planet-host and non-host stars.  To accomplish this objective, we performed a bootstrapped Monte Carlo simulation.  The process is described below.  Note that we first removed the three planet-hosts and five non-planet-hosts at [Fe/H]~$<$~-0.2~dex from our sample, as this region is very poorly populated.
\\
\\
\underline{Create realization of planet hosts}

\begin{enumerate}

\item We created a realization of the observed planet hosts, as follows:

\begin{itemize}

\item We randomly drew a number between 30 and 60, to determine the size of the realization.

\item We then selected this number of planet hosts, using random sampling with replacement (meaning some stars may have been duplicates).

\end{itemize}

\suspend{enumerate}
\underline{Calculate [Fe/H] histogram for planet hosts}
\resume{enumerate}

\item We determined the [Fe/H] distribution of the planet-host realization by calculating a histogram with bins of width 0.1~dex.

\suspend{enumerate}
\underline{Create realization of non-hosts}
\resume{enumerate}

\item We created a realization of the observed non-host stars by randomly selecting -- again with replacement -- a number of stars equal to the size of the host realization, while forcing the [Fe/H] distribution to match (or as closely as possible) that of the host sample.  This was done in the following manner:

\begin{itemize}

\item We randomly drew a number between zero and one.  

\item If this probability was lower than the normalized [Fe/H] distribution of the host set at the metallicity of the non-host, we included the star in our non-host realization.  If not, we rejected the selection.

\item This process was repeated until a number of non-hosts equal to the size of the host realization was selected.

\end{itemize}

\suspend{enumerate}
\underline{Calculate [Fe/H] histogram for non-hosts}
\resume{enumerate}

\item We determined the [Fe/H] distribution of the non-host realization by calculating a histogram with bins of width 0.1 dex.

\suspend{enumerate}
\underline{Evaluate difference}
\resume{enumerate}

\item We then performed two-sided Kolmogorov-Smirnov (K-S) tests on the [Fe/H], [Si/Fe] and [O/Fe] distributions of the two sets.

\item The entire process was repeated $10^{6}$ times, and the K-S probability was noted for each trial.

\end{enumerate}

The process described in step~3 above, in which we construct pairs of samples with iron abundance distributions that match as closely as possible, is crucial to our experiment, since we do not wish to simply reproduce the known planet-metallicity correlation.  Rather, we wish to know whether differences exist between our host and control sets at a given [Fe/H].  Since any minor differences in the distributions of a single trial will average out over $10^{6}$ trials, forcing the iron distributions to match in this manner serves as a method for binning our data by [Fe/H].

Given the finite nature of our samples, the statistical procedure explained above offers an excellent method for determining any possible compositional differences in individual elements between host and non-host stars.  The results of our experiment are explained in the following section.

\section{Results}
\label{results}

Our statistical investigation of the silicon and oxygen content of stars hosting giant planets reveals a distinct compositional difference for silicon, but not for oxygen, when these stars are compared to similarly iron-rich non-host stars.  In figure 8 we show that the [Si/Fe] distributions of planet-hosts and non-hosts are significantly different.  The variable on the horizontal axis is the K-S probability, which is a measure of the statistical significance of the difference between the cumulative distributions of two samples.  In this sense, it measures the probability that two samples were drawn from the same parent distribution.   Thus, a low K-S probability is consistent with the host and non-host samples being drawn from different distributions, while a high probability indicates similar parent distributions.  The peak at a K-S probability of zero in figure 8 is strong evidence of such a difference.  In figure 9 we show the same plot for oxygen, where we see little evidence of a difference for [O/Fe].

The dashed line in figures 8 and 9 depicts the K-S probability for [Fe/H], which shows that our matching of host stars to equally iron-rich non-host stars was not perfect.  For perfect control sets in each trial, we would expect the dashed lines to have single peaks at probability~=~1.  Better matching could be achieved with smaller histogram bin widths, but this would require unrealistically larger samples to draw from.  We note, however, that we have devised a mathematical method to control for this imperfect matching, as discussed below.



In order to quantify the difference, or lack thereof, in the distributions of Si and O, we devised a ``total'' probability P, representing the product of integrated K-S probabilities for [Fe/H] and [Si/Fe] or [O/Fe] divided by the integrated [Fe/H] probability squared: 
\begin{equation}
P_{X} = \frac{\int_{0}^{1} f_{[X/Fe]}f_{[Fe/H]}pdp}{\int_{0}^{1} f_{[Fe/H]}^{2}pdp}
\label{probability}
\end{equation}
where \emph{f} represents the percentage in a particular probability bin \emph{p} of width \emph{dp}.  This equation represents a method of controlling for spurious low K-S statistics that result from the [Fe/H] distributions of our two sets not matching perfectly in some trials.  That is, if the underlying Fe distributions don't match, we can't expect the Si or O distributions to match.  

With this definition in hand, and using a probability bin size of dp~=~0.10, we find a total probability for Si and O of:
\begin{equation}
P_{Si} = 0.01
\label{silicon}
\end{equation}
\begin{equation}
P_{O} = 0.57
\label{oxygen}
\end{equation}
The small total Si probability is consistent with the hosts and non-hosts in our sample being drawn from separate parent distributions of stars.  Put differently, there is only a 1\% chance that the planet-harboring stars and non-planet-harboring stars in our sample were drawn from the same parent distribution.  The results of our statistical analysis therefore suggest a significant difference in the Si abundances of planet host stars, when compared to stars hosting no known giant planets.  The rather large total O probability is consistent with the samples being drawn from the same parent distribution (a 57\% chance).

\section{Conclusions and Future Work}
\label{conclusions}

We have determined stellar atmospheric parameters and derived differential abundances of Fe, Si, and O for a uniform sample of 76 planet-host stars and 80 non-host stars, using high resolution and high signal-to-noise data obtained on the 2.7m and HET telescopes at McDonald Observatory.  We find a statistically significant difference in the [Si/Fe] distribution between the two groups of stars.  This result lends strong support to the core-accretion theory of planet formation, since much of the solid material available for core formation is thought to consist of silicate grains with icy mantles.  We find no statistically significant difference in the [O/Fe] distributions.  Although the uncertainties in our oxygen measurements are significantly larger than for silicon, we nevertheless find it unlikely that a statistically significant trend -- at the level of that seen with silicon -- would emerge with smaller error bars.  Reducing the error bars would likely require 3D model atmospheres and the incorporation of non-LTE effects in the line formation process.  Such models are becoming available (Asplund 2005), and future studies could likely measure oxygen to higher precision with the use of these.

The lack of a trend with oxygen is a surprising result, as we would expect this alpha element to track the silicon abundance (as predicted in Robinson et al. 2006).  Our interpretation is that the stellar photospheres are tracing species important for grain nucleation, rather than subsequent icy mantle growth.  Since silicon rather than oxygen is the limiting reagent for grain nucleation, the entire process of dust formation would in that case depend on the silicon abundance.  Oxygen is so over-abundant relative to refractory species that the process of core accretion may be insensitive to variations in the overall oxygen abundance.  We liken the process to cloud formation on Earth, in which condensation nuclei play a key role.  The atmosphere may be virtually saturated with water vapor, but without seeds (such as dust, sea salt and bacteria) onto which this vapor can condense, no clouds can form.  We posit that silicon and other refractory elements serve as these seeds in protostellar discs.  Without them, the process of giant planet formation may be independent of the amount of volatile material available.

When comparing our results to the work of others, we note the studies by Mel\'{e}ndez et al. (2009) and Ram\'{i}rez et al. (2009, 2010), who argue that the Sun is deficient in refractory elements relative to volatile elements when compared to nearby solar twins and solar analogs.  They attribute this difference to efficient planet formation around the Sun, whereby refractory elements were preferentially locked up in the terrestrial planets during the early protostellar period.  At first glance, this result seems to contradict our findings, but further inspection reveals that for silicon specifically the results are inconclusive.  Mel\'{e}ndez et al. (2009) actually find that Si is enhanced by $\backsim$0.03~dex in the Sun when compared to the average Si abundance of their sample of 21 solar twins and analogs, while Ram\'{i}rez et al. (2009) find a $\backsim$0.03~dex decrement in the Sun when compared to the average Si abundance of their sample of 64 solar twins and analogs.  Ram\'{i}rez et al. (2010) report no difference at all in their meta-analysis of solar analogs from six independent studies.  Further studies addressing any possible difference in the silicon content of the Sun compared to solar twins and analogs need to be performed before any definite conclusions can be made. 

Based on our results and interpretation, we predict that planet detection rate should correlate positively with host star abundance for those elements responsible for grain nucleation, and that no such trend should exist for the most abundant volatile elements responsible for icy mantle growth.  The most important refractory contributors to the composition of dust in planet-forming regions are thought to be silicon, iron, magnesium, sulfur and aluminum; while the most important volatiles are probably oxygen, carbon and nitrogen.  Carbon is an interesting case in that it might contribute significant mass both to grain nuclei and icy mantles.  Henning \& Salama (1998) argue that up to 20\% of the carbon in the universe is probably locked in refractory grains, while simulations by Dodson-Robinson \& Bodenheimer (2010) demonstrate that the ice giants Uranus and Neptune required solid methane in their feeding zones to grow to their present size.  Hence, we expect it to contribute significant amounts of mass to giant planet cores; and we predict that planet detection rate should correlate positively with host star carbon abundance for any population of planets formed by core accretion.

The present analysis represents an important ``first look'' study in which we focused on the single most abundant volatile contributor to dust grains (oxygen), and the single most important refractory contributor (silicon) after iron.  Our future work will involve a more comprehensive analysis, in which we will increase our sample sizes, to allow for better matching of the overall metallicity distributions of our planet-host and control samples, and increase the number of studied species to include the most abundant elements (discussed above) thought to be important for planet-formation.

\vspace{5 mm}

E.B.'s and S.D.R.'s work was funded by the University of Texas through a faculty startup package awarded to S.D.R.  W.D.C.'s work was made possible by funding through NASA Origins grant NNX09AB30G.  C.S.'s work was made possible by funding through NSF grant AST~09-08978.

This research has made use of the NASA/IPAC/NExScI Star and Exoplanet Database, which is operated by the Jet Propulsion Laboratory, California Institute of Technology, under contract with the National Aeronautics and Space Administration.  This research has also made use of the SIMBAD database, operated at CDS, Strasbourg, France.  The Hobby-Eberly Telescope is a joint project of the University of Texas at Austin, the Pennsylvania State University, Stanford University, the Ludwig Maximilians-Universit\"{a}t M\"{u}nchen, and the Georg-August-Universit\"{a}t at G\"{o}ttingen. The HET is named in honor of its principal benefactors, William P. Hobby and Robert E. Eberly.

The authors thank Ian Roederer for the use of his IDL equivalent width measurement software; Ivan Ram\'{i}rez for providing his grid of non-LTE oxygen corrections, along with an IDL routine for interpolating within this grid; and Julia Bryson for help measuring iron abundances.   We are grateful to the anonymous referee, whose careful reading and many suggestions helped improve the manuscript.

Finally, we would like to thank all the people who have helped gather data over the years for the 2.7m radial velocity search program at McDonald Observatory, including Phillip MacQueen, Mike Endl, Paul Robertson, Rob Wittenmyer, Diane Paulsen, and Artie Hatzes.

\begin{center}

\begin{deluxetable}{lccccccc}															
\tabletypesize{\scriptsize}															
\tablecaption{Summary of Results for Planet-Host Stars \label{results}}															
\tablehead{\colhead{Star Name} & \colhead{[Fe/H]} & \colhead{[Si/Fe]} & \colhead{[O/Fe]} & \colhead{[O/Fe]} & \colhead{$T_{\rm eff}$} & \colhead{$\log \: g$} & \colhead{$v_{\rm mic}$ } \\ & & & {(LTE)\tablenotemark{a}} & {(NLTE)\tablenotemark{a}} & {(K)} & & {(km s$^{-1}$)}}															
\startdata															
109 Psc	&	0.19	&	-0.04	&	-0.08	&	-0.10	&	5675	&	4.12	&	1.16	\\
14 Her	&	0.51	&	0.03	&	-0.28	&	-0.19	&	5355	&	4.47	&	1.07	\\
16 Cyg B\tablenotemark{$\ast$}	&	0.06	&	0.01	&	0.05	&	0.03	&	5705	&	4.36	&	1.13	\\
47 UMa\tablenotemark{$\ast\ast$}	&	0.05	&	0.01	&	0.01	&	-0.02	&	5880	&	4.40	&	1.16	\\
51 Peg	&	0.25	&	0.02	&	-0.04	&	-0.03	&	5800	&	4.50	&	1.03	\\
55 Cnc	&	0.38	&	0.10	&	-0.10	&	-0.04	&	5250	&	4.49	&	1.11	\\
6 Lyn	&	-0.04	&	0.02	&	0.15	&	0.07	&	4990	&	3.34	&	1.26	\\
61 Vir	&	0.03	&	0.00	&	0.08	&	0.08	&	5550	&	4.42	&	1.00	\\
70 Vir\tablenotemark{$\ast\ast$}	&	-0.01	&	-0.02	&	0.03	&	0.01	&	5549	&	4.14	&	1.18	\\
eps Eri	&	-0.02	&	-0.04	&	0.14	&	0.19	&	5110	&	4.54	&	1.11	\\
HD 100777	&	0.33	&	0.07	&	-0.09	&	-0.05	&	5585	&	4.44	&	0.98	\\
HD 102195	&	0.11	&	0.01	&	0.05	&	0.10	&	5270	&	4.56	&	1.13	\\
HD 106252	&	-0.05	&	-0.01	&	0.11	&	0.06	&	5870	&	4.41	&	1.07	\\
HD 107148	&	0.33	&	0.07	&	-0.04	&	-0.02	&	5810	&	4.56	&	1.08	\\
HD 114762	&	-0.67	&	0.16	&	0.43	&	0.33	&	5960	&	4.54	&	1.17	\\
HD 118203	&	0.15	&	0.09	&	0.26	&	0.14	&	5690	&	3.87	&	1.15	\\
HD 11964	&	0.14	&	-0.02	&	-0.06	&	-0.07	&	5345	&	4.02	&	1.18	\\
HD 12661	&	0.39	&	0.04	&	-0.07	&	-0.04	&	5720	&	4.42	&	1.22	\\
HD 130322	&	0.09	&	-0.03	&	-0.09	&	-0.05	&	5410	&	4.48	&	1.12	\\
HD 132406	&	0.16	&	0.00	&	-0.04	&	-0.03	&	5820	&	4.48	&	1.01	\\
HD 136118	&	-0.02	&	-0.04	&	0.27	&	0.10	&	6095	&	4.07	&	1.12	\\
HD 136418	&	-0.04	&	0.05	&	0.01	&	-0.02	&	4985	&	3.50	&	1.03	\\
HD 13931	&	0.10	&	-0.04	&	0.07	&	0.02	&	5850	&	4.26	&	1.14	\\
HD 1461	&	0.23	&	0.01	&	-0.07	&	-0.05	&	5745	&	4.51	&	1.19	\\
HD 149026	&	0.31	&	0.04	&	0.08	&	0.02	&	6140	&	4.35	&	1.23	\\
HD 149143	&	0.25	&	0.07	&	0.15	&	0.07	&	5825	&	4.05	&	1.15	\\
HD 154345	&	-0.08	&	0.02	&	0.06	&	0.09	&	5430	&	4.54	&	0.75	\\
HD 155358	&	-0.61	&	0.10	&	0.55	&	0.41	&	5860	&	4.24	&	0.75	\\
HD 16175	&	0.36	&	0.05	&	0.18	&	0.14	&	6020	&	4.39	&	1.28	\\
HD 164922	&	0.21	&	0.09	&	-0.03	&	0.02	&	5395	&	4.57	&	0.90	\\
HD 168443	&	0.12	&	0.01	&	0.18	&	0.15	&	5580	&	4.22	&	1.17	\\
HD 178911 B	&	0.14	&	-0.01	&	0.29	&	0.17	&	5730	&	3.97	&	1.18	\\
HD 185269	&	0.13	&	0.04	&	0.19	&	0.07	&	5990	&	4.03	&	1.26	\\
HD 189733	&	0.01	&	0.07	&	0.13	&	0.18	&	5020	&	4.55	&	0.82	\\
HD 190228	&	-0.20	&	-0.02	&	0.03	&	0.00	&	5310	&	3.91	&	1.22	\\
HD 195019	&	0.07	&	-0.04	&	0.10	&	0.05	&	5790	&	4.24	&	1.26	\\
HD 19994	&	0.19	&	-0.01	&	0.22	&	0.08	&	6095	&	4.05	&	1.32	\\
HD 202206	&	0.36	&	-0.04	&	-0.22	&	-0.18	&	5770	&	4.50	&	1.15	\\
HD 20367	&	0.14	&	-0.08	&	0.11	&	0.06	&	6120	&	4.51	&	1.18	\\
HD 20782	&	0.01	&	-0.07	&	0.07	&	0.05	&	5770	&	4.45	&	1.12	\\
HD 209458	&	0.01	&	0.03	&	0.18	&	0.09	&	6090	&	4.40	&	1.17	\\
HD 210277	&	0.28	&	0.04	&	0.01	&	0.04	&	5565	&	4.51	&	1.04	\\
HD 217107	&	0.45	&	-0.03	&	-0.20	&	-0.14	&	5690	&	4.55	&	1.13	\\
HD 219828	&	0.25	&	-0.01	&	0.03	&	-0.01	&	5895	&	4.25	&	1.18	\\
HD 30562	&	0.24	&	0.00	&	0.10	&	0.03	&	5860	&	4.13	&	1.25	\\
HD 33283	&	0.36	&	0.02	&	0.14	&	0.06	&	5995	&	4.16	&	1.39	\\
HD 34445	&	0.20	&	-0.05	&	0.12	&	0.07	&	5830	&	4.24	&	1.13	\\
HD 3651	&	0.17	&	0.06	&	0.04	&	0.08	&	5185	&	4.38	&	1.10	\\
HD 37124	&	-0.41	&	0.20	&	0.49	&	0.47	&	5505	&	4.57	&	0.87	\\
HD 38529	&	0.40	&	-0.03	&	-0.03	&	-0.07	&	5600	&	3.90	&	1.40	\\
HD 40979	&	0.23	&	-0.02	&	0.10	&	0.04	&	6160	&	4.42	&	1.10	\\
HD 43691	&	0.31	&	0.03	&	0.09	&	0.02	&	6225	&	4.33	&	1.19	\\
HD 44219	&	0.04	&	0.01	&	0.16	&	0.10	&	5710	&	4.21	&	1.31	\\
HD 45350	&	0.33	&	0.01	&	-0.04	&	-0.02	&	5605	&	4.35	&	1.15	\\
HD 45652	&	0.33	&	0.07	&	-0.07	&	-0.01	&	5340	&	4.52	&	0.83	\\
HD 46375	&	0.30	&	0.08	&	-0.01	&	0.05	&	5250	&	4.51	&	1.04	\\
HD 49674	&	0.34	&	0.07	&	-0.05	&	0.00	&	5630	&	4.61	&	0.93	\\
HD 50554	&	-0.04	&	0.01	&	0.27	&	0.18	&	5915	&	4.33	&	1.12	\\
HD 52265	&	0.21	&	0.02	&	0.23	&	0.15	&	6105	&	4.38	&	1.34	\\
HD 60532	&	-0.06	&	-0.01	&	0.43	&	0.17	&	6220	&	3.88	&	1.18	\\
HD 66428	&	0.34	&	0.07	&	-0.03	&	0.01	&	5765	&	4.62	&	1.11	\\
HD 6718	&	0.03	&	-0.04	&	-0.01	&	0.00	&	5745	&	4.53	&	0.98	\\
HD 68988	&	0.36	&	0.07	&	0.01	&	0.02	&	5960	&	4.56	&	1.10	\\
HD 72659	&	0.01	&	-0.03	&	0.16	&	0.06	&	5870	&	4.16	&	1.23	\\
HD 73534	&	0.23	&	0.08	&	-0.01	&	0.00	&	4965	&	3.71	&	1.08	\\
HD 75898	&	0.20	&	0.05	&	0.23	&	0.11	&	5880	&	4.01	&	1.24	\\
HD 81040	&	-0.06	&	-0.01	&	0.07	&	0.08	&	5730	&	4.60	&	0.80	\\
HD 82943	&	0.30	&	-0.02	&	0.03	&	0.02	&	5975	&	4.47	&	1.20	\\
HD 8574	&	-0.04	&	0.00	&	0.21	&	0.09	&	6010	&	4.22	&	1.35	\\
HD 88133	&	0.41	&	0.04	&	-0.11	&	-0.09	&	5475	&	4.16	&	1.12	\\
HD 89307	&	-0.14	&	0.02	&	0.24	&	0.18	&	5915	&	4.47	&	1.21	\\
HD 92788	&	0.37	&	0.02	&	-0.07	&	-0.02	&	5800	&	4.61	&	1.06	\\
HD 9446	&	0.14	&	-0.02	&	0.07	&	0.07	&	5770	&	4.55	&	1.20	\\
HD 96167	&	0.36	&	0.07	&	0.06	&	0.02	&	5775	&	4.14	&	1.22	\\
HIP 14810	&	0.27	&	0.10	&	0.05	&	0.06	&	5510	&	4.30	&	1.08	\\
rho CrB	&	-0.18	&	0.03	&	0.23	&	0.17	&	5825	&	4.37	&	1.02	\\
\enddata
\tablenotetext{a}{For the ``NLTE'' [O/Fe] values we incorporated the non-LTE corrections of Ram\'{i}rez, Allende Prieto \& Lambert (2007).  The ``LTE'' values are our original abundance determinations, ignoring any possible non-LTE effects (see section 3.3).}
\tablenotetext{$\ast$}{This star was used to estimate our systematic uncertainties, by analyzing the sensitivity of our derived abundances to variations in the stellar atmospheric parameters (see section 4).}													
\tablenotetext{$\ast\ast$}{These stars were used to estimate our random uncertainties, by analyzing multiple observations (see section 4).}											
\end{deluxetable}

\clearpage

\begin{deluxetable}{lccccccc}															
\tabletypesize{\scriptsize}															
\tablecaption{Summary of Results for Non-Host Stars \label{results}}															
\tablehead{\colhead{Star Name} & \colhead{[Fe/H]} & \colhead{[Si/Fe]} & \colhead{[O/Fe]} & \colhead{[O/Fe]} & \colhead{$T_{\rm eff}$} & \colhead{$\log \: g$} & \colhead{$v_{\rm mic}$ } \\ & & & {(LTE)\tablenotemark{a}} & {(NLTE)\tablenotemark{a}} & {(K)} & & {(km s$^{-1}$)}}															
\startdata															
10 CVn	&	-0.43	&	-0.01	&	0.18	&	0.14	&	5900	&	4.57	&	0.85	\\
11 Aqr	&	0.27	&	0.01	&	0.05	&	0.02	&	5905	&	4.30	&	1.21	\\
13 Ori	&	-0.16	&	0.05	&	0.26	&	0.18	&	5740	&	4.25	&	1.20	\\
13 Tri	&	-0.10	&	-0.03	&	0.11	&	0.01	&	5950	&	4.18	&	1.17	\\
18 Cet	&	-0.18	&	-0.03	&	0.14	&	0.07	&	5840	&	4.17	&	1.30	\\
31 Aql	&	0.46	&	0.02	&	-0.05	&	-0.02	&	5635	&	4.34	&	1.21	\\
36 UMa	&	-0.02	&	-0.09	&	0.06	&	-0.01	&	6150	&	4.42	&	1.00	\\
58 Eri	&	0.04	&	-0.02	&	0.02	&	0.02	&	5830	&	4.58	&	1.10	\\
83 Leo A	&	0.38	&	0.04	&	-0.20	&	-0.13	&	5472	&	4.50	&	1.06	\\
88 Leo A	&	0.03	&	-0.03	&	0.06	&	0.02	&	6000	&	4.50	&	1.12	\\
alp For	&	-0.14	&	-0.04	&	0.17	&	0.02	&	6250	&	4.17	&	1.20	\\
beta Com	&	0.10	&	-0.04	&	-0.02	&	-0.04	&	6060	&	4.56	&	1.06	\\
gam Lep B	&	0.11	&	-0.06	&	-0.19	&	-0.11	&	4990	&	4.61	&	1.20	\\
gam2 Del	&	0.31	&	-0.11	&	-0.26	&	-0.27	&	4935	&	3.15	&	1.53	\\
HD 10086	&	0.13	&	-0.04	&	-0.01	&	0.01	&	5670	&	4.52	&	1.18	\\
HD 105844	&	0.33	&	-0.03	&	-0.19	&	-0.14	&	5590	&	4.48	&	0.98	\\
HD 107146	&	0.00	&	-0.06	&	0.10	&	0.08	&	5870	&	4.56	&	1.18	\\
HD 108942	&	0.28	&	-0.02	&	0.02	&	0.00	&	5770	&	4.23	&	1.28	\\
HD 110010	&	0.38	&	0.00	&	0.10	&	0.10	&	6010	&	4.52	&	1.28	\\
HD 11007	&	-0.17	&	0.00	&	0.14	&	0.04	&	6015	&	4.24	&	1.27	\\
HD 110537	&	0.12	&	-0.02	&	0.05	&	0.04	&	5690	&	4.35	&	1.30	\\
HD 111431	&	0.09	&	0.00	&	0.20	&	0.10	&	5880	&	4.13	&	1.27	\\
HD 115043	&	0.01	&	-0.05	&	0.03	&	0.01	&	5840	&	4.47	&	0.99	\\
HD 116956	&	0.11	&	0.01	&	0.04	&	0.07	&	5325	&	4.41	&	1.21	\\
HD 129357	&	0.02	&	-0.02	&	0.10	&	0.06	&	5750	&	4.32	&	1.17	\\
HD 13825	&	0.22	&	0.02	&	-0.01	&	-0.01	&	5660	&	4.35	&	1.26	\\
HD 138776	&	0.44	&	-0.02	&	-0.13	&	-0.12	&	5700	&	4.25	&	1.18	\\
HD 149028	&	0.21	&	0.00	&	-0.05	&	-0.05	&	5520	&	4.22	&	1.23	\\
HD 184385	&	0.13	&	0.01	&	-0.02	&	0.02	&	5565	&	4.61	&	1.24	\\
HD 184499	&	-0.40	&	0.13	&	0.48	&	0.40	&	5830	&	4.50	&	0.96	\\
HD 185414	&	-0.10	&	-0.01	&	0.06	&	0.04	&	5820	&	4.55	&	1.23	\\
HD 187748	&	0.08	&	-0.04	&	0.13	&	0.08	&	5980	&	4.44	&	1.18	\\
HD 190613	&	0.04	&	-0.01	&	0.15	&	0.10	&	5720	&	4.22	&	0.91	\\
HD 19256	&	0.25	&	-0.01	&	0.11	&	0.04	&	5910	&	4.14	&	1.33	\\
HD 200078	&	0.25	&	0.03	&	0.24	&	0.19	&	5630	&	4.14	&	1.28	\\
HD 221146	&	0.12	&	0.04	&	0.11	&	0.05	&	5880	&	4.30	&	1.24	\\
HD 299	&	0.20	&	-0.05	&	0.07	&	0.02	&	6000	&	4.35	&	1.22	\\
HD 31609	&	0.26	&	-0.06	&	-0.08	&	-0.03	&	5560	&	4.50	&	1.08	\\
HD 39480	&	0.19	&	0.00	&	0.20	&	0.11	&	5750	&	4.00	&	1.24	\\
HD 47127	&	0.14	&	0.02	&	0.08	&	0.08	&	5615	&	4.43	&	1.15	\\
HD 56124	&	0.00	&	-0.02	&	0.07	&	0.04	&	5750	&	4.35	&	1.12	\\
HD 59062	&	0.38	&	0.03	&	-0.06	&	-0.03	&	5575	&	4.37	&	1.04	\\
HD 60521	&	0.13	&	0.00	&	0.18	&	0.12	&	5805	&	4.22	&	1.25	\\
HD 73350	&	0.13	&	-0.01	&	0.01	&	0.01	&	5815	&	4.57	&	1.23	\\
HD 75880	&	0.16	&	0.04	&	0.08	&	0.06	&	5595	&	4.25	&	1.23	\\
HD 8038	&	0.17	&	-0.01	&	0.19	&	0.17	&	5590	&	4.32	&	1.18	\\
HD 87000	&	0.14	&	0.01	&	-0.06	&	0.00	&	5170	&	4.49	&	1.16	\\
HD 92719	&	-0.04	&	-0.07	&	0.04	&	0.02	&	5760	&	4.42	&	0.94	\\
HD 94126	&	0.40	&	0.07	&	-0.12	&	-0.09	&	5570	&	4.30	&	0.97	\\
HD 94482	&	-0.02	&	-0.05	&	0.15	&	0.04	&	5995	&	4.15	&	1.33	\\
HD 95653	&	0.54	&	-0.04	&	-0.26	&	-0.20	&	5585	&	4.35	&	0.93	\\
HD 97037	&	-0.05	&	-0.03	&	0.12	&	0.07	&	5830	&	4.32	&	1.18	\\
HD 97854	&	0.20	&	-0.03	&	0.10	&	0.00	&	5985	&	4.06	&	1.38	\\
HD 99505	&	-0.11	&	-0.07	&	0.09	&	0.08	&	5700	&	4.48	&	0.93	\\
HR 173	&	-0.56	&	0.20	&	0.49	&	0.42	&	5360	&	4.09	&	1.01	\\
HR 1980	&	0.06	&	-0.04	&	0.03	&	-0.01	&	6085	&	4.53	&	1.17	\\
HR 2208	&	-0.01	&	-0.05	&	-0.02	&	0.00	&	5700	&	4.55	&	1.24	\\
HR 2225	&	0.02	&	-0.04	&	-0.05	&	-0.03	&	5590	&	4.52	&	1.17	\\
HR 2721	&	-0.30	&	0.01	&	0.25	&	0.18	&	5860	&	4.40	&	1.04	\\
HR 2997	&	-0.06	&	-0.03	&	-0.02	&	0.01	&	5470	&	4.52	&	1.10	\\
HR 3538	&	0.13	&	0.01	&	0.01	&	0.02	&	5775	&	4.57	&	1.08	\\
HR 3862	&	-0.02	&	-0.04	&	0.16	&	0.06	&	6180	&	4.41	&	1.18	\\
HR 3881	&	0.14	&	-0.01	&	0.15	&	0.07	&	5915	&	4.20	&	1.24	\\
HR 4051	&	0.05	&	-0.04	&	0.13	&	0.03	&	6040	&	4.29	&	1.26	\\
HR 448	&	0.18	&	-0.01	&	0.08	&	0.01	&	5840	&	4.07	&	1.50	\\
HR 4525	&	-0.18	&	-0.01	&	0.09	&	0.11	&	5600	&	4.59	&	0.98	\\
HR 4767	&	-0.05	&	-0.03	&	0.10	&	0.05	&	6010	&	4.52	&	1.04	\\
HR 4864	&	0.14	&	-0.03	&	-0.08	&	-0.04	&	5630	&	4.57	&	1.15	\\
HR 5183	&	0.07	&	0.00	&	0.15	&	0.07	&	5810	&	4.15	&	1.32	\\
HR 6669	&	0.08	&	-0.06	&	0.08	&	-0.03	&	6140	&	4.24	&	1.12	\\
HR 7569	&	-0.13	&	0.07	&	0.32	&	0.25	&	5720	&	4.31	&	1.18	\\
HR 8964	&	0.14	&	-0.05	&	-0.05	&	-0.04	&	5840	&	4.57	&	1.27	\\
iota Psc	&	-0.05	&	-0.02	&	0.19	&	0.05	&	6240	&	4.24	&	1.16	\\
kap1 Cet	&	0.07	&	-0.01	&	0.00	&	0.01	&	5705	&	4.51	&	1.11	\\
lam Aur	&	0.13	&	-0.02	&	0.06	&	0.02	&	5899	&	4.34	&	1.10	\\
lam Ser	&	0.03	&	-0.01	&	0.06	&	0.00	&	5920	&	4.25	&	1.22	\\
mu Her	&	0.34	&	-0.05	&	-0.07	&	-0.08	&	5600	&	4.06	&	1.35	\\
pi1 UMa	&	-0.03	&	-0.03	&	0.09	&	0.06	&	5820	&	4.49	&	1.14	\\
tau Cet	&	-0.44	&	0.11	&	0.27	&	0.29	&	5345	&	4.54	&	0.54	\\
xi Boo A	&	-0.09	&	-0.03	&	-0.01	&	0.03	&	5530	&	4.63	&	1.20	\\
\tablenotetext{a}{For the ``NLTE'' [O/Fe] values we incorporated the non-LTE corrections of Ram\'{i}rez, Allende Prieto \& Lambert (2007).  The ``LTE'' values are our original abundance determinations, ignoring any possible non-LTE effects (see section 3.3).}
\enddata															
\end{deluxetable}

\clearpage

\begin{deluxetable}{cccc}
\tabletypesize{\scriptsize}
\tablecaption{List of Fe I lines\label{iron1linelist}
}
\tablehead{ \colhead{Wavelength} & \colhead{Excitation Potential} 
& \colhead{Oscillator Strength} & \colhead{Solar Equivalent Width} \\ {(Angstroms)} & {(eV)} & {(log \emph{gf})} & {(m\r{A})}}
\startdata
4445.47	&	0.09	&	-5.38	&	44.5	\\
4537.67	&	3.27	&	-2.88	&	19.6	\\
4556.93	&	3.25	&	-2.69	&	28.2	\\
4593.54	&	3.94	&	-2.06	&	30.4	\\
4788.75	&	3.24	&	-1.76	&	69.5	\\
4873.75	&	3.30	&	-3.06	&	12.8	\\
5123.72	&	1.01	&	-3.06	&	116.6	\\
5127.68	&	0.05	&	-6.12	&	22.5	\\
5151.91	&	1.01	&	-3.32	&	105.1	\\
5213.81	&	3.94	&	-2.76	&	6.5	\\
5247.05	&	0.09	&	-4.98	&	72.4	\\
5250.21	&	0.12	&	-4.90	&	75.4	\\
5295.30	&	4.42	&	-1.69	&	28.4	\\
5373.70	&	4.47	&	-0.87	&	65.6	\\
5386.34	&	4.15	&	-1.77	&	32.9	\\
5560.21	&	4.43	&	-1.19	&	51.9	\\
5577.03	&	5.03	&	-1.55	&	13.0	\\
5636.70	&	3.64	&	-2.61	&	21.4	\\
5705.47	&	4.30	&	-1.60	&	38.5	\\
5753.12	&	4.26	&	-0.69	&	87.6	\\
5778.45	&	2.59	&	-3.59	&	21.6	\\
5811.92	&	4.14	&	-2.43	&	10.6	\\
5814.81	&	4.28	&	-1.97	&	22.1	\\
5849.68	&	3.69	&	-2.99	&	7.5	\\
5858.78	&	4.22	&	-2.26	&	13.2	\\
5927.79	&	4.65	&	-1.09	&	44.3	\\
5956.69	&	0.86	&	-4.50	&	57.6	\\
6034.03	&	4.31	&	-2.42	&	8.8	\\
6120.24	&	0.92	&	-5.95	&	5.6	\\
6151.62	&	2.18	&	-3.37	&	51.2	\\
6159.37	&	4.61	&	-1.97	&	11.7	\\
6165.36	&	4.14	&	-1.47	&	46.2	\\
6187.99	&	3.94	&	-1.72	&	48.5	\\
6226.73	&	3.88	&	-2.20	&	29.8	\\
6265.13	&	2.18	&	-2.54	&	92.5	\\
6380.75	&	4.19	&	-1.38	&	55.5	\\
6392.54	&	2.28	&	-4.03	&	17.8	\\
6498.94	&	0.96	&	-4.69	&	46.7	\\
6509.61	&	4.08	&	-2.98	&	3.6	\\
6591.33	&	4.59	&	-2.06	&	10.5	\\
6593.87	&	2.43	&	-2.37	&	90.7	\\
6597.56	&	4.79	&	-1.06	&	42.8	\\
6608.02	&	2.28	&	-4.04	&	18.3	\\
6609.11	&	2.56	&	-2.66	&	72.5	\\
6646.93	&	2.61	&	-3.99	&	11.0	\\
6667.42	&	2.45	&	-4.40	&	5.6	\\
6667.73	&	4.58	&	-2.15	&	9.6	\\
6699.16	&	4.59	&	-2.18	&	8.5	\\
6703.57	&	2.76	&	-3.15	&	37.4	\\
6704.48	&	4.22	&	-2.66	&	5.7	\\
6710.32	&	1.49	&	-4.87	&	16.1	\\
6725.35	&	4.10	&	-2.30	&	17.6	\\
6732.07	&	4.58	&	-2.21	&	6.8	\\
6739.52	&	1.56	&	-4.94	&	11.0	\\
6745.09	&	4.58	&	-2.17	&	9.1	\\
6745.95	&	4.08	&	-2.76	&	6.3	\\
6746.95	&	2.61	&	-4.35	&	4.8	\\
6753.47	&	4.56	&	-2.28	&	6.2	\\
6837.02	&	4.59	&	-1.80	&	17.7	\\
6839.83	&	2.56	&	-3.45	&	32.3	\\
6843.65	&	4.55	&	-0.93	&	65.5	\\
6851.63	&	1.61	&	-5.31	&	5.4	\\
6857.24	&	4.08	&	-2.16	&	22.4	\\
6862.49	&	4.56	&	-1.57	&	30.7	\\
6978.85	&	2.48	&	-2.45	&	88.3	\\
\enddata
\end{deluxetable}

\clearpage

\begin{deluxetable}{cccccl}
\tabletypesize{\scriptsize}
\tablecaption{List of Fe II lines\label{iron2linelist}
}
\tablehead{ \colhead{Wavelength} & \colhead{Excitation Potential} 
& \colhead{Oscillator Strength}  & \colhead{Solar Equivalent Width} \\ {(Angstroms)} & {(eV)} & {(log \emph{gf})} & {(m\r{A})}}
\startdata
4413.60	&	2.68	&	-3.79	&	39.9	\\
4491.40	&	2.86	&	-2.71	&	80.0	\\
4582.84	&	2.84	&	-3.18	&	61.7	\\
4620.52	&	2.83	&	-3.21	&	62.9	\\
5132.67	&	2.81	&	-4.08	&	24.9	\\
5197.58	&	3.23	&	-2.22	&	89.6	\\
5234.62	&	3.22	&	-2.18	&	91.4	\\
5264.81	&	3.23	&	-3.13	&	48.0	\\
5325.55	&	3.22	&	-3.16	&	48.2	\\
5414.07	&	3.22	&	-3.58	&	31.7	\\
6084.11	&	3.20	&	-3.79	&	21.1	\\
6149.26	&	3.89	&	-2.69	&	39.6	\\
6247.56	&	3.89	&	-2.30	&	57.3	\\
6369.46	&	2.89	&	-4.11	&	20.7	\\
6383.72	&	5.55	&	-2.24	&	9.4	\\
6416.92	&	3.89	&	-2.64	&	43.2	\\
6446.41	&	6.22	&	-1.97	&	4.5	\\
6516.08	&	2.89	&	-3.31	&	62.1	\\
7222.39	&	3.89	&	-3.26	&	19.6	\\
7224.49	&	3.89	&	-3.20	&	24.9	\\
7515.83	&	3.90	&	-3.39	&	14.8	\\
7711.72	&	3.90	&	-2.50	&	53.5	\\
\enddata
\end{deluxetable}

\begin{figure}
\plotone{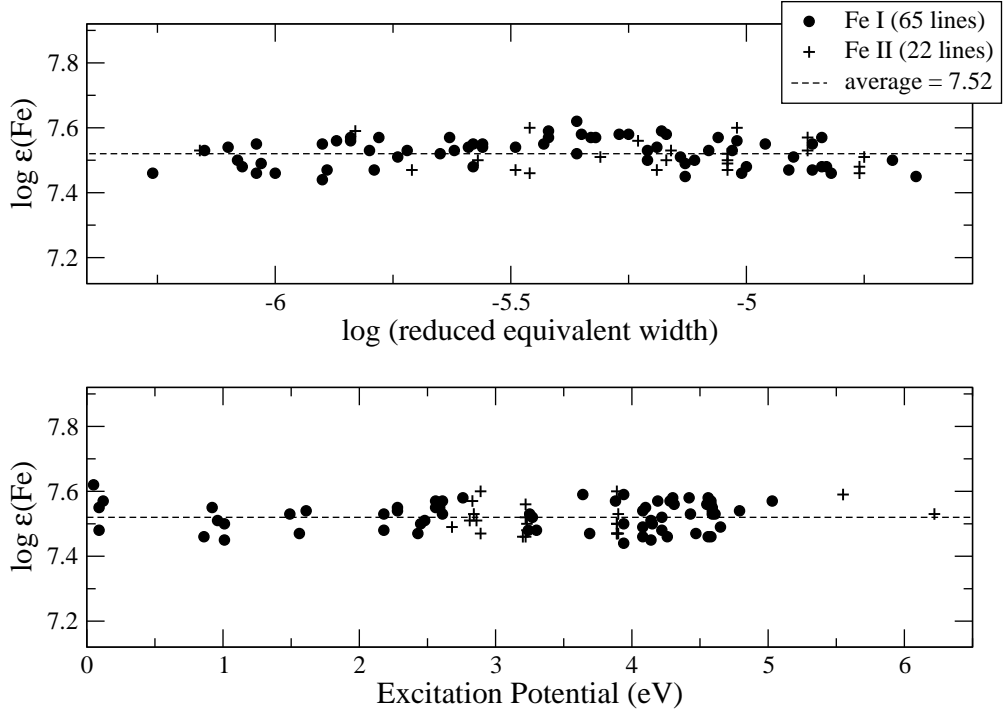}
\caption{Plots of $\log \: \epsilon$(Fe) for each measured iron line in the Sun.  A similar analysis was performed on each of our target stars in order to determine stellar atmospheric parameters.  Effective temperature was constrained by eliminating any trend in iron abundance with excitation potential; microturbulent velocity by eliminating any trend with reduced equivalent width; and surface gravity by forcing the derived abundances of Fe I and Fe II to match.  The top panel shows the derived solar iron abundance as a function of reduced equivalent width (=~EW/$\lambda$); the bottom panel as a function of excitation potential.  Fe I is represented by filled circles; Fe II by crosses.}
\label{solport}
\end{figure}

\begin{deluxetable}{cccccl}
\tabletypesize{\scriptsize}
\tablecaption{List of Si I lines\label{silicon1linelist}}
\tablehead{ \colhead{Wavelength} & \colhead{Excitation Potential} 
& \colhead{Oscillator Strength}  \\ {(Angstroms)} & {(eV)} & {(log \emph{gf})} }
\startdata
5645.61	&	4.93	&	-2.10	\\
5665.56	&	4.92	&	-2.07	\\
5684.48	&	4.95	&	-1.62	\\
5708.40	&	4.95	&	-1.47	\\
5772.15	&	5.08	&	-1.71	\\
5793.07	&	4.93	&	-2.05	\\
\enddata
\end{deluxetable}

\begin{figure}
\plotone{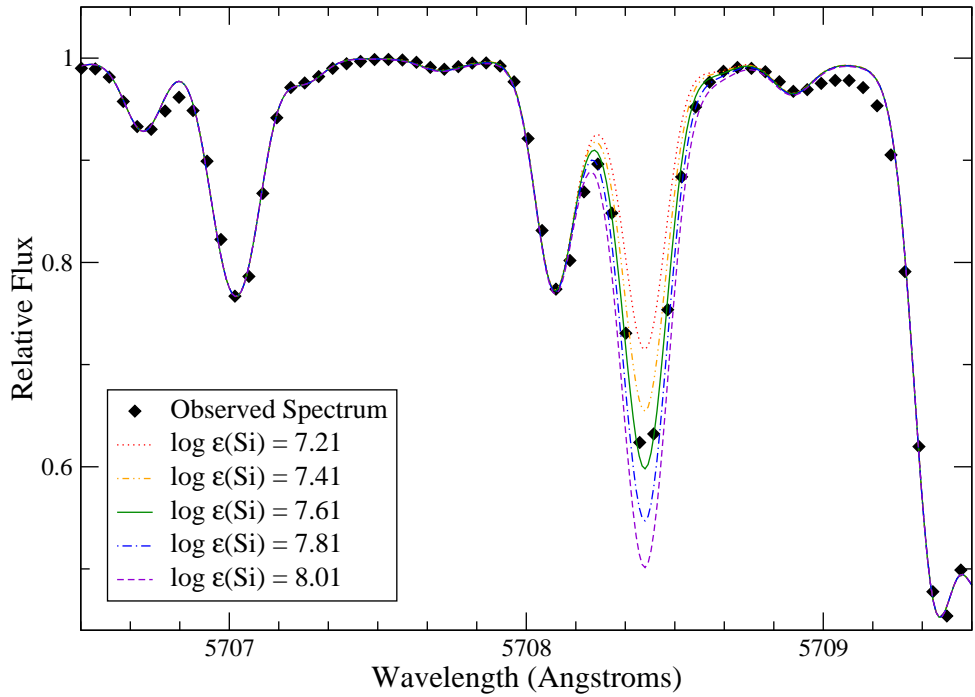}
\caption{A MOOG-synthesized portion of the solar spectrum around the 5708 \r{A} Si I absorption feature.  Similar synthetic spectra were used to determine Si and O abundances in each of our target stars, by minimizing the residuals to the fits of the various observed spectra.  The plot displays the observed solar port spectrum as diamond symbols, and the synthetic spectra as lines.  In this synthesis, the silicon abundance was varied by $\pm$ 0.2~dex and $\pm$~0.4 dex from the best-fit value of $\log \: \epsilon$(Si)~=~7.61.}
\label{siliconsynth}
\end{figure}



\clearpage

\begin{figure}
\plotone{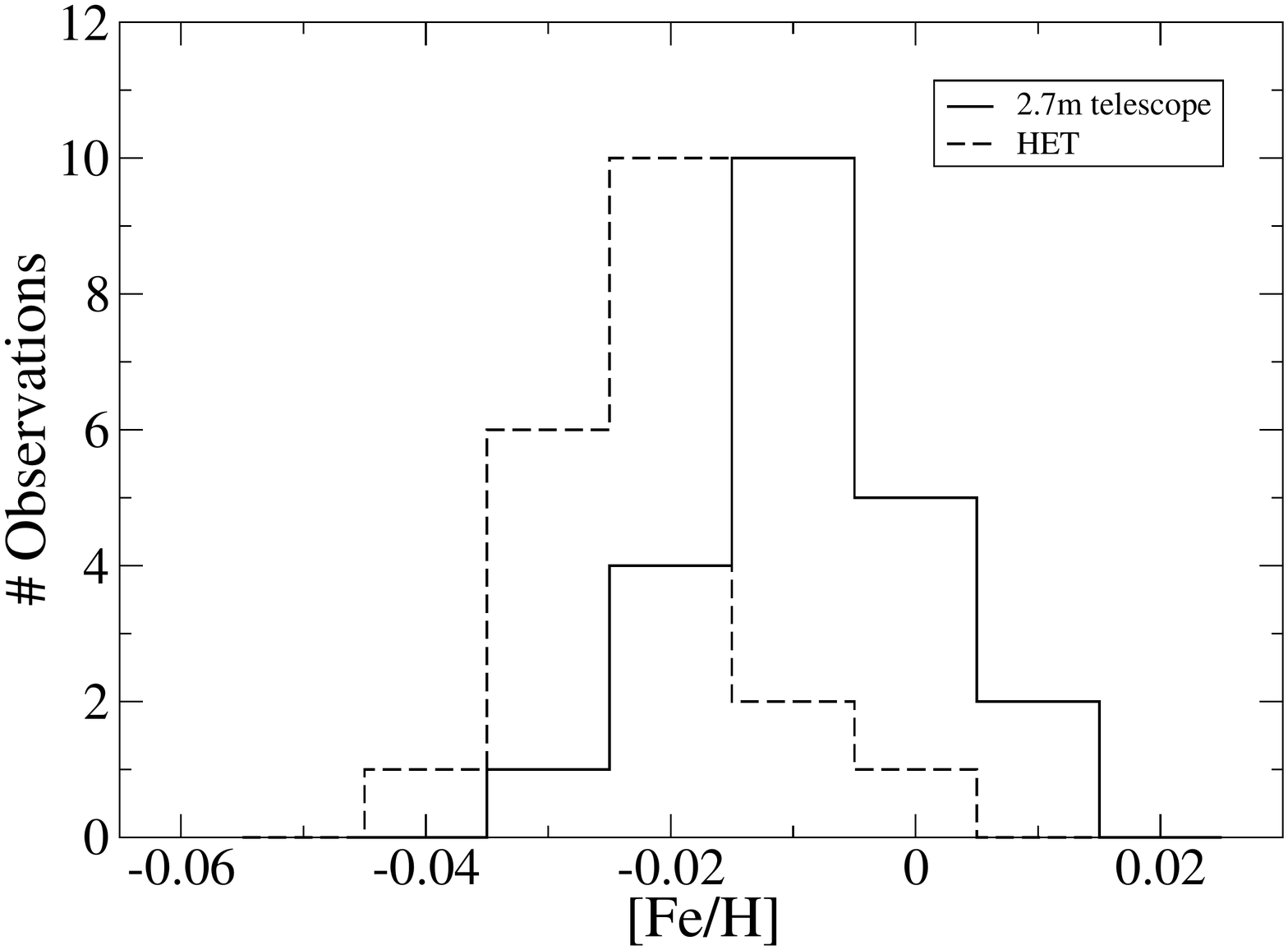}
\caption{Histograms showing the number of observations of 70 Vir as a function of [Fe/H] for the 2.7m telescope (solid line, 22 observations in total) and HET (dashed line, 20 observations in total).  The distributions appear roughly gaussian, with a standard deviation of 0.01 dex in both cases.  Note the small bin size of 0.01 dex.  Our measurements are highly repeatable, with a minimal offset of 0.01 dex in the mean derived abundances from the two instruments.  This difference is well within our error bars and is likely due to the differing number of pixels per resolution element on the two detectors.}
\label{70Vir_FeH}
\end{figure}

\begin{figure}
\plotone{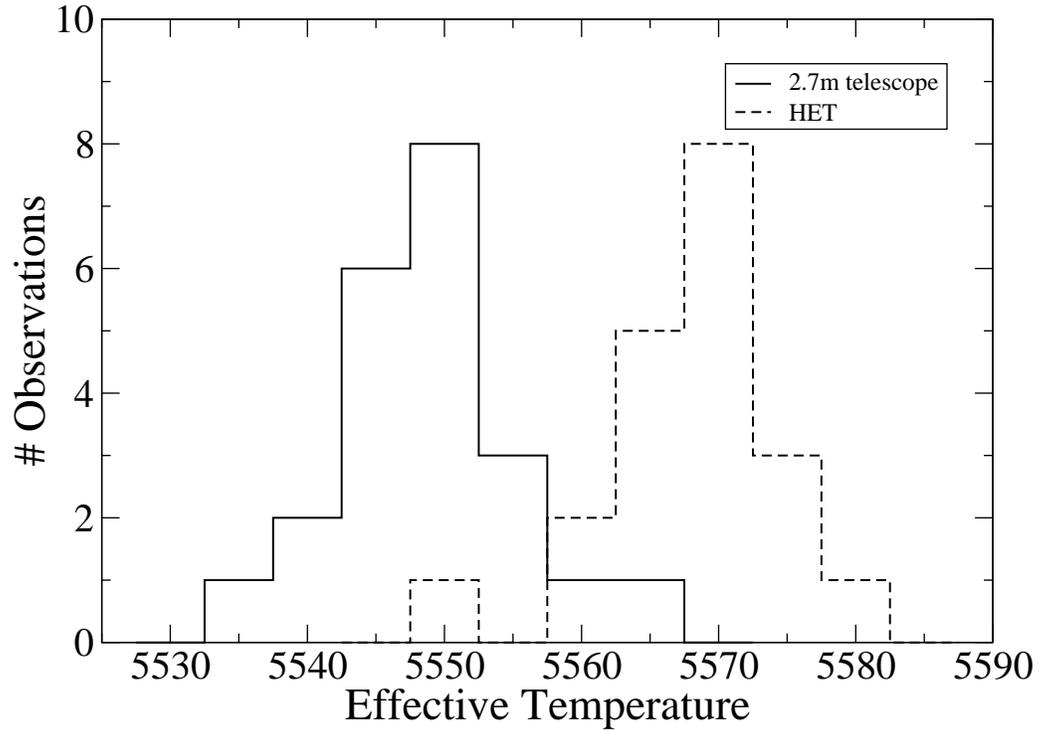}
\caption{Histograms showing the number of observations of 70 Vir as a function of effective temperature for the 2.7m telescope (solid line, 22 observations in total) and HET (dashed line, 20 observations in total).  The distributions appear roughly gaussian, with a standard deviation of 10 K in both cases.  Note the small bin size of 5 K.  Our measurements are highly repeatable, with a minimal offset of 20 K in the mean derived temperatures from the two instruments.  This difference is well within our error bars and is likely due to the differing number of pixels per resolution element on the two detectors.}
\label{70Vir_Teff}
\end{figure}

\begin{figure}
\plotone{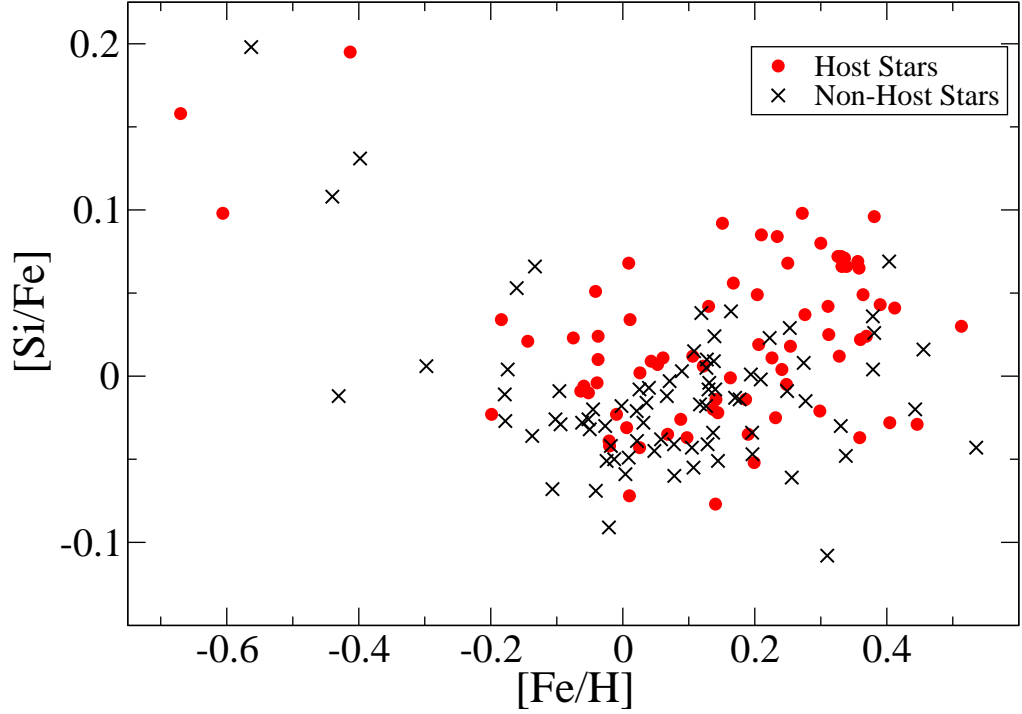}
\caption{A plot of [Si/Fe] as a function of [Fe/H] for our sample.  At iron abundances above [Fe/H] of -0.2~dex, where the vast majority of our sample lies, we observe that the planet-hosting stars in our sample tend to be enhanced in silicon when compared to stars without any known giant planets.  Two-thirds of the host stars lie at or above [Si/Fe] of zero, while three-fourths of our non-host stars lie at or below [Si/Fe] of zero.  We note that the distribution agrees well with galactic chemical evolution models and observations (e.g. Timmes et al. 1995).  Planet-hosting stars are represented by filled circles; non-host stars by crosses.}
\label{silicon}
\end{figure}

\begin{figure}
\plotone{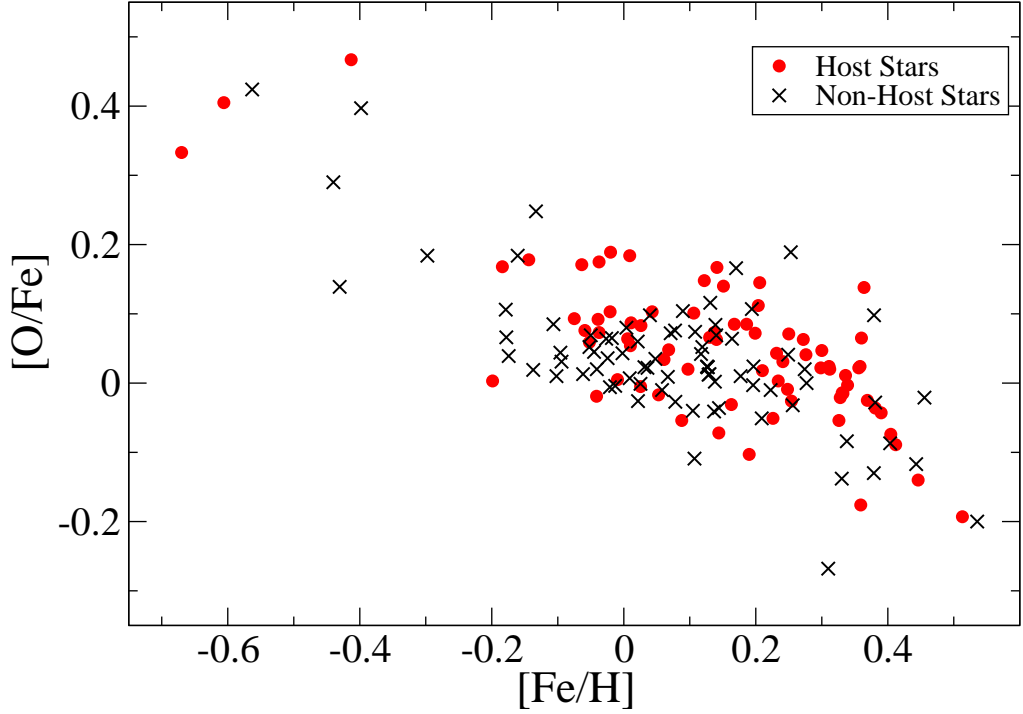}
\caption{A plot of [O/Fe] as a function of [Fe/H] for our sample.  No discernible trends are apparent between planet-hosting stars and non-host stars in our sample.  We note that the distribution agrees well with galactic chemical evolution models and observations (e.g. Timmes et al. 1995).  Planet-hosts are represented by filled circles; non-hosts by crosses.}
\label{oxygen}
\end{figure}

\begin{figure}
\plotone{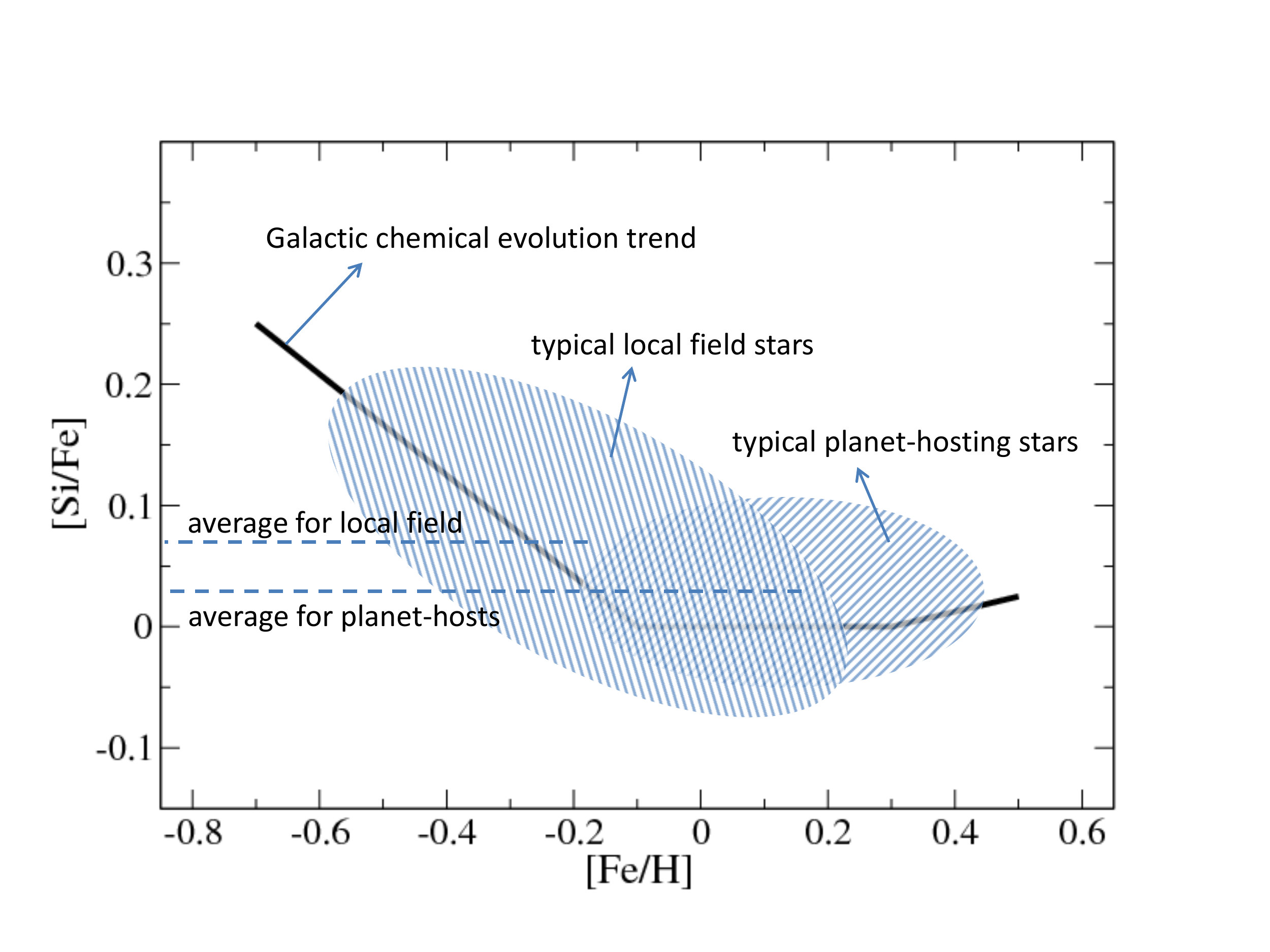}
\caption{A cartoon demonstrating the need for careful matching of the underlying [Fe/H] distributions for any sample of planet-hosting stars to the control set of field stars.  Planet hosts appear to follow the same general galactic chemical evolution trend as typical field stars, but tend to preferentially lie on the iron-rich end of the distribution (Bodaghee et al. 2003, Santos et a. 2005).  Since the trend is not flat, the expected average [Si/Fe] for the typical local field is different than the expected average for planet-hosting stars.  It is therefore imperative that the underlying [Fe/H] distribution of the control sample match that of the planet-host sample, by selecting local field stars that are more metal-rich than average.}
\label{cartoon}
\end{figure}



\begin{figure}
\begin{turn}{180}
\plotone{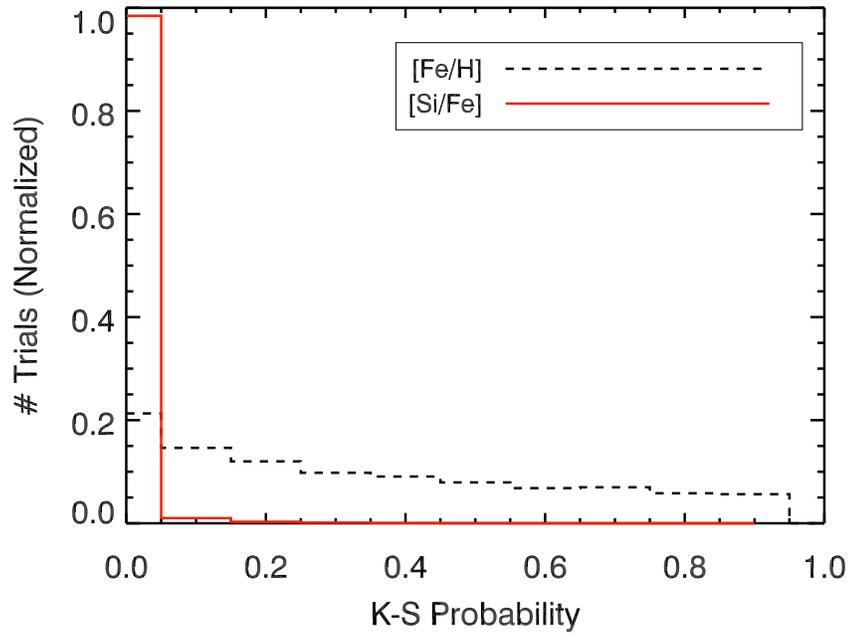}
\end{turn}
\caption{A histogram of the percentage of trials vs. probability for [Si/Fe] (solid line) and [Fe/H] (dashed line; shown for reference), with bin widths of 0.10.  The peak at a K-S probability of zero is strong evidence of a difference in the Si abundances of our planet-host and non-host samples.}
\label{KS_SiFe}
\end{figure}

\begin{figure}
\begin{turn}{180}
\plotone{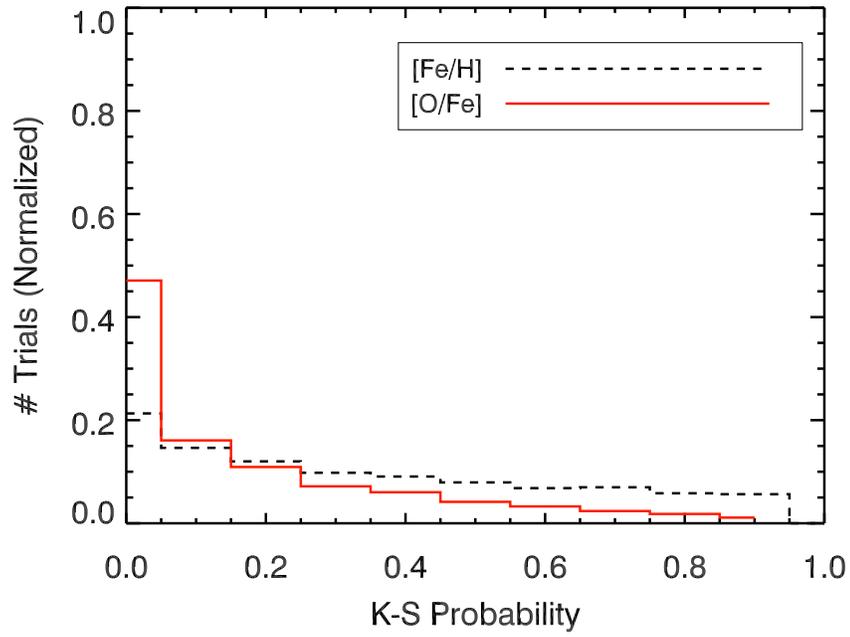}
\end{turn}
\caption{A histogram of the percentage of trials vs. probability for [O/Fe] (solid line) and [Fe/H] (dashed line; shown for reference), with bin widths of 0.10.  The histogram for [O/Fe] appears qualitatively similar to that for [Fe/H], indicating no significant difference in the O abundances of our planet-host and non-host samples.}
\label{KS_OFe}
\end{figure}

\end{center}


\begin{thebibliography}{9999}

\bibitem[Asplund(2005)]{Asplund05} Asplund, M. 2005, \araa, 43, 481
\bibitem[Asplund(2009)]{Asplund09} Asplund, M., Grevesse, N., Sauval, J., \& Scott, P. 2009, \araa, 47, 481
\bibitem[Bodaghee(2003)]{bodaghee03} Bodaghee, A., Santos, N.~C., Israelian, G. \& Mayor, M. 2003, \aap, 404, 715
\bibitem[Dodson-Robinson(2010)]{dodson-robinson10} Dodson-Robinson, S.~E. \& Bodenheimer, P. 2010, Icarus, 207, 491
\bibitem[Fabbian(2009)]{fabbian09} Fabbian, D., Asplund, M., Barklem, P.~S., Carlsson, M., \& Kiselman, D. 2009, \aap, 500, 1221
\bibitem[Fabbian(2010)]{fabbian10} Fabbian, D., Khomenko, E., Moreno-Insertis, F., \& Nordlund, \AA. 2010, \apj, 724, 1536
\bibitem[Fischer(2005)]{fischer05} Fischer, D.~A., \& Valenti, J. 2005, \apj, 622, 1102
\bibitem[Fuhrmann(2008)]{fuhrmann08} Fuhrmann, K., \& Bernkopf, J. 2008, \mnras, 384, 1563
\bibitem[Gonzalez(1997)]{gonzalez97} Gonzalez, G. 1997, \mnras, 285, 403
\bibitem[Gonzalez(1998)]{gonzalez98} Gonzalez, G. 1998, \aap, 334, 221
\bibitem[Gonzalez(1999)]{gonzalez99} Gonzalez, G. 1999, \mnras, 308, 447
\bibitem[Hayashi(1981)]{hayashi81} Hayashi, C. 1981, Supplement of the Progress of Theoretical Physics, No. 70, 35
\bibitem[Henning(1998)]{henning98} Henning, Th. \& Salama, F. 1998, Science, 282, 2204
\bibitem[Kiselman(1993)]{kiselman93} Kiselman, D. 1993, \aap, 275, 269
\bibitem[Kiselman(2001)]{kiselman01} Kiselman, D. 2001, New Astron. Rev., 45, 559
\bibitem[Kurucz(1993)]{kurucz93a} Kurucz, R. 1993, ATLAS9 Stellar Atmosphere Programs and 2 km/s grid. Kurucz CD-ROM No. 13. (Cambridge: Smithsonian Astrophys. Obs.)
\bibitem[Kurucz(1993)]{kurucz93b} Kurucz, R. 1993, Atomic data for opacity calculations. Kurucz CD-ROM No. 1. (Cambridge: Smithsonian Astrophys. Obs.)
\bibitem[Lodders(2004)]{lodders04} Lodders, K. 2004, \apj, 611, 587
\bibitem[Mel\'{e}ndez(2009)]{melendez09a} Mel\'{e}ndez, J., \& Barbuy, B. 2009, \aap, 497, 611
\bibitem[Mel\'{e}ndez(2009)]{melendez09b} Mel\'{e}ndez, J., Asplund, M., Gustafsson, B., \& Yong, D. 2009, \apj, 704, L66
\bibitem[Neves(2009)]{neves09} Neves, V., Santos, N.~C., Sousa, S.~G., Correia, A.~C.~M., \& Israelian, G. 2009, \aap, 497, 563
\bibitem[O'Brian(1991)]{obrian91} O'Brian, T.~R., Wickliffe, M.~E., Lawler, J.~E., Whaling, W., \& Brault, J.~W. 1991, J. Opt. Soc. Am., B8, 1185
\bibitem[Pollack(1996)]{pollack96} Pollack, J.~B., Hubickyj, O., Bodenheimer, P., Lissauer, J.~J., Podolak, M., \& Greenzweig, Y. 1996, Icarus, 124, 62
\bibitem[Ram\'{i}rez(2007)]{ramirez07} Ram\'{i}rez, I., Allende~Prieto, C., \& Lambert, D.~L. 2007, \aap, 465, 271
\bibitem[Ram\'{i}rez(2009)]{ramirez09} Ram\'{i}rez, I., Mel\'{e}ndez, J., \& Asplund, M. 2009, \aap, 508, L17
\bibitem[[Ram\'{i}rez(2010)]{ramirez10} Ram\'{i}rez, I., Asplund, M., Baumann, P., Mel\'{e}ndez, J., \& Bensby, T. 2010, \aap, 521, A33
\bibitem[Roederer(2010)]{roederer10} Roederer, I.~U., Sneden, C., Thompson, I.~B., Preston, G.~W., \& Shectman, S. 2010, \apj, 711, 573
\bibitem[Robinson(2006)]{robinson06} Robinson, S.~E., Laughlin, G., Bodenheimer, P., \& Fischer, D. 2006, \apj, 643, 484
\bibitem[Safronov(1969)]{safronov69} Safronov, V.~S. 1969, Evolution of the Protoplanetary Cloud and the Formation of the Earth and Planets. Nauka, Moscow. English translation: NASA TTF-667, 1972
\bibitem[Santos(2005)]{santos05} Santos, N.~C., Israelian, G., Mayor, M., Bento, J.~P., Almeida, P.~C., Sousa, S.~G., \& Ecuvillon, A. 2005, \aap, 437, 1127
\bibitem[Shi(2008)]{shi08} Shi, J.~R., Gehren, T., Butler, K., Mashonkina, L.~I., \& Zhao, G. 2008, \aap, 486, 303
\bibitem[Sneden(1973)]{sneden73} Sneden, C.~A. 1973, Ph.D. thesis, Univ. of Texas at Austin
\bibitem[Takeda(2005)]{takeda05} Takeda, Y., \& Honda, S. 2005, \pasj, 57, 65 
\bibitem[Timmes(1995)]{timmes95} Timmes, F.~X., Woosley, S.~E., \& Weaver, T.~A. 1995, \apjs, 98, 617
\bibitem[Tull(1994)]{tull94} Tull, R.~G., MacQueen, P., Sneden, C., \& Lambert, D.~L. 1994, in ASP Conf. Ser. 55, Optical Astronomy from the Earth and Moon, ed. D.~M. Pyper \& R.~J. Angione (San Fransisco: ASP), 148
\bibitem[Tull(1998)]{tull98} Tull, R.~G. 1998, in Optical Astronomical Instrumentation, Proc. SPIE 3355, 387
\bibitem[Wedemeyer(2001)]{wedemeyer01} Wedemeyer, S. 2001, \aap, 373, 998
\bibitem[Weidenschilling(1977)]{weidenschilling77} Weidenschilling, S.~J. 1977, \apss, 51, 153
\bibitem[Wittenmyer(2006)]{wittenmyer06} Wittenmyer, R.~A., Endl, M., Cochran, W.~D., Hatzes, A.~P., Walker, G.~A.~H., Yang, S.~L.~S., \& Paulson, D.~P. 2006, \aj, 132, 177

\end{thebibliography}
\end{document}